\documentclass[11pt]{article}

\usepackage{amsthm}

\usepackage[T1]{fontenc}
\usepackage[utf8]{inputenc}
\usepackage{newtxtext, newtxmath}
\usepackage[protrusion=true, expansion=true]{microtype}
\usepackage{setspace}

\usepackage[english]{babel}
\usepackage{csquotes}

\usepackage[margin=1in, columnsep=20pt]{geometry}

\usepackage{booktabs}
\usepackage[hang, small, labelfont=bf, textfont=it, up]{caption}
\usepackage{enumitem}
\setlist[itemize]{noitemsep}
\usepackage{graphicx}
\usepackage{longtable}
\usepackage{ragged2e}
\usepackage{rotating}
\usepackage{threeparttable}

\usepackage{abstract}

\usepackage{titlesec}
\titleformat*{\section}{\large\bfseries}
\titleformat*{\subsection}{\bfseries}

\usepackage[hidelinks]{hyperref}
\usepackage[authordate, backend=biber]{biblatex-chicago}
\usepackage{titling}

\title{
	\Large\bfseries
	A Classifier-Lasso Approach for Estimating Production Functions with Latent Group Structures\thanks{I would like to thank Amrei Stammann, Joel Stiebale, and participants and organizers of the 2021 International Panel Data Conference (IPDC).}
}
\author{
	\normalsize
	Daniel Czarnowske\thanks{
		Heinrich-Heine-Universität, Chair of Statistics and Econometrics, Universitätsstr. 1, 40225 Düsseldorf, Germany, phone: +49 211 81-10620, e-mail: \texttt{\href{mailto:daniel.czarnowske@hhu.de}{daniel.czarnowske@hhu.de}}
	}
}
\date{\normalsize\today}

\DeclareMathOperator{\argmin}{\arg\,\min\;}

\DeclareMathOperator{\eye}{\mathbf{1}}
\DeclareMathOperator{\iid}{\text{iid.}\;}
\DeclareMathOperator{\ind}{\mathbf{1}}

\DeclareMathOperator{\EX}{\mathbb{E}}

\DeclareMathOperator{\N}{\mathcal{N}}

\newcommand{\norm}[1]{\lVert #1 \rVert}
\theoremstyle{definition}
\newtheorem{algorithm}{Algorithm}

\newtheorem{example}{Example}
\newtheorem{remark}{Remark}

\begin{document}

% Title page with abstract
\maketitle
\begin{abstract}
	\noindent I present a new estimation procedure for production functions with latent group structures. I consider production functions that are heterogeneous across groups but time-homogeneous within groups, and where the group membership of the firms is unknown. My estimation procedure is fully data-driven and embeds recent identification strategies from the production function literature into the classifier-Lasso. Simulation experiments demonstrate that firms are assigned to their correct latent group with probability close to one. I apply my estimation procedure to a panel of Chilean firms and find sizable differences in the estimates compared to the standard approach of classification by industry.
	\vfill
\end{abstract}
\thispagestyle{empty}
\clearpage

% Main text
\onehalfspacing
\setcounter{page}{1}

\section{Introduction}
\label{sec:introduction}

Production functions are one of the central concepts in economic theory, with origins dating back at least to the mid-18th century (see \cite{h1997} for the history of production functions). It is all the more remarkable that the identification of production functions is still an econometric challenge.\footnote{\textcites{gnr2020}{s2020} are two very recent examples that solve this identification problem for gross output production functions.} The main challenge is an endogeneity problem first mentioned by \textcite{ma1944}. It arises when firms make input decisions based on their productivity, while productivity is usually unobservable to the researcher. Since the seminal work of \textcites{op1996}{lp2003}, all recent identification strategies, such as those of \textcites{bb2000}{acf2015}{gnr2020}, use panel data along with assumptions about firm behavior to solve the endogeneity problem. 

Although these identification strategies are frequently applied in practice, a remaining challenge is latent firm heterogeneity in the data. For instance, firms may differ in their production technologies, which can be thought of model parameter heterogeneity. Therefore, it is common practice to classify firms based on prior information, like an industry classification, and then estimate separate production functions. I will refer to this approach as \textit{ex-ante classification} throughout the paper. However, as pointed out by \textcite{gm1999}, even in a narrowly defined industry where firms produce nearly identical products, these firms may use their inputs very differently, which in turn may indicate different production technologies. On the other hand, firms in different industries may nevertheless use very similar production technologies so that neglecting this similarity reduces the precision of the estimation. 

To overcome these limitations, I propose a new estimation procedure. Using examples of three identification strategies, I show how to embed them into the classifier-Lasso (\textit{C-Lasso}) of \textcite{ssp2016}. My estimation procedure has no additional data requirements, except that the panel must be sufficiently long, and can even be used in situations where prior information is not available at all. I demonstrate good finite-sample performance of my estimator in simulation experiments. Even in moderately long panels, my estimation procedure is able to assign all firms to the correct latent group with probabilities approaching one. The properties of my estimator are asymptotically identical to an infeasible estimator that knows and exploits the latent group structure. Further, because the iterative estimation algorithm proposed by \textcite{ssp2016} is computationally intensive or even infeasible for panels with many firms, I present an extension to the algorithm that overcomes this problem. Finally, I apply my estimation procedure to a panel of Chilean firms and find that an ex-ante classification by industry does not match the \textit{data-driven classification} of my estimator. In particular, I classify firms from five industries into three latent groups, where each of the five industries is composed of firms from all three latent groups. A simple comparison of the mean squared residuals, reveals that the estimation based on the data-driven classification fits the data better than the ex-post classification, although the latter estimates more parameters. Moreover, estimating separate production functions by industry is not only less efficient, but also yields significantly different estimates and thus leads to different conclusions.

Besides the C-Lasso of \textcite{ssp2016}, there are at least two alternative approaches to deal with latent group structures. The first approach is a finite mixture specification, which models the probability of belonging to a latent group as a function of the explanatory variables. Some applications of parametric and semi-parametric finite mixtures are \textcites{s2005}{ks2009}{bc2014}. The second approach is a modification of the k-means clustering algorithm. Examples are \textcites{ln2012}{bm2015}{sw2015}. Of the two alternative approaches, k-means clustering is probably the most popular. \textcite[][Remark 1]{swj2019} name and discuss three major differences between C-Lasso and k-means clustering: i) C-Lasso has an additional tuning parameter, ii) k-means forces group assignment, and iii) k-means is computationally more demanding. There are pros and cons for i) and ii), so the main argument for C-Lasso is iii), which is particularly relevant for applications with many firms due to the large number of potential cluster partitions that have to be tried by k-means. Both alternative approaches have been recently adapted to the estimation of production functions. \textcite{kss2017} use a finite mixture specification to extend the identification strategy of \textcite{gnr2020}. They also provide empirical evidence of additional firm heterogeneity within narrowly defined industries in a panel of Japanese firms. \textcite{css2019} extend the k-means clustering algorithm to account for multidimensional heterogeneity and reassess the rise of markups reported by \textcite{dleu2020}. They find that the level and growth of markups are lower when additional firm heterogeneity within industries is taken into account. I contribute to this literature by presenting an alternative estimation procedure based on the C-Lasso and providing additional empirical evidence that classification by industry is not sufficient to fully account for firm heterogeneity.

The paper is organized as follows. I introduce production functions with latent group structures in Section \ref{sec:model}. I show how these production functions can be estimated in Section \ref{sec:estimation}. I analyze the performance of my estimation procedure in Section \ref{sec:simulations}. I provide an empirical illustration using Chilean panel data in Section \ref{sec:illustration}. Finally, I give some concluding remarks in Section \ref{sec:conclusion}.

Throughout this paper, I follow conventional notation: scalars are represented in standard type, vectors and matrices in boldface, and all vectors are column vectors. Further, $\norm{\cdot}$ denotes the Euclidean norm.

\section{Production Functions with Latent Group Structures}
\label{sec:model}

\subsection{Structural Panel Model}
\label{sec:panel_model}

% Data requirements
I consider panel data of $N$ firms observed for $T + 1$ periods $\{\xi_{it} \colon i \in \{1, \ldots, N\}, t \in \{0, \ldots, T\}\}$, where $\xi_{it} \coloneqq (Y_{it}, \mathbf{X}_{it}, \mathbf{Z}_{it})$ is a collection of variables for firm $i$ at time $t$. $Y_{it}$ is the output, $\mathbf{X}_{it}$ is a vector of inputs, like capital $K_{it}$, labor $L_{it}$, or intermediate inputs $M_{it}$, and $\mathbf{Z}_{it}$ is a vector of additional firm characteristics, like input and output prices or a firm's export status. I follow standard convention and denote the corresponding natural log values in lowercase letters, i.e. $y_{it}$ is the natural logarithm of $Y_{it}$.

% Base model
To allow for heterogeneous production technologies, I follow \textcites{ln2012}{bm2015}{ssp2016} and assume a time-homogeneous latent group structure where each firm belongs to exactly one of $J^{0}$ groups. I collect all firms that are members of a latent group $j \in \{1, \ldots, J^{0}\}$ in a set $G_{j}^{0}$, where $\bigcup_{j = 1}^{J^{0}} G_{j}^{0} = \{1, \ldots, N\}$ and $G_{j}^{0} \cap G_{j^{\prime}}^{0} = \varnothing$ for all $j \neq j^{\prime}$. I assume that $J^{0}$ is known to the researcher, but the group membership of each firm is not. All firms in $G_{j}^{0}$ have the same well-defined production function
\begin{equation}
	\label{eq:basemodel}
	y_{it} \coloneq P(\mathbf{x}_{it}) \boldsymbol{\beta}_{i}^{0} + \omega_{it} + \epsilon_{it} \, ,
\end{equation}
where $P(\mathbf{x}_{it})$ is a basis system of transformed input variables, $\boldsymbol{\beta}_{i}^{0} \coloneqq \sum_{j = 1}^{J^{0}} \ind\{i \in G_{j}^{0}\} \boldsymbol{\alpha}_{j}^{0}$ are the corresponding model parameters that follow a general group pattern, $\omega_{it}$ is a predictable productivity shock, unobservable to the researcher, and $\epsilon_{it}$ is an unanticipated productivity shock or a measurement error. Further, I assume that $\norm{\boldsymbol{\alpha}_{j}^{0} - \boldsymbol{\alpha}_{j^{\prime}}^{0}} > 0$ for all $j \neq j^{\prime}$, i.e. firms in different latent groups use different production technologies.

% Econometric issues
Besides the unknown latent group membership of the firms, the estimation of \eqref{eq:basemodel} poses further econometric challenges. Some important issues are related to the measurements of output and inputs, the functional form of the production function, and endogeneity problems (see \cite{abbp2007} and \cite{a2019} for more details).\footnote{It is known since \textcite{ma1944} that when firms choose their inputs optimally, at least to some extent, $\omega_{it}$ and $\mathbf{x}_{it}$ are correlated, leading to inconsistent estimators of output elasticities (see \cite{gm1999}).} In particular, these problems led to very different estimation strategies that can be readily applied if the researcher knows the latent group membership of the firms.\footnote{For instance, it is common practice to classify firms by an industry classification.} Traditional identification strategies rely on instrumental variables or fixed effects. Both strategies are not very popular in recent work, because the former requires additional excluded instruments, like output and input prices, and the latter requires the restrictive assumption $\omega_{it} = \omega_{i}$.\footnote{These traditional strategies can already be applied using estimators presented in Section 2 and 3 of \textcite{ssp2016}.} Modern identification strategies are based on modeling firm behavior in a dynamic environment. The key assumptions are related to the timing of firms' input decisions and the information available to firms at the time of these decisions. There are three popular estimation strategies: control function / proxy variable approach (\cite{op1996}, \cite{lp2003}, and \cite{acf2015}), dynamic panel estimator (\cite{ab1991}, \cite{bb1998}, and \cite{bb2000}), and first-order condition approach (\cite{m1987} and \cite{gnr2020}). All these approaches derive moment conditions from the underlying structural model and use these moments to estimate the model parameters by the generalized method of moments (GMM).

\subsection{Moment Conditions for Three Modern Identification Strategies}
\label{sec:firm_moments}

% Restrictions
Because there are numerous modifications and extensions of modern estimation strategies, I focus on the three most popular baseline strategies. In particular, I briefly summarize the control function approach of \textcite{acf2015}, the dynamic panel estimator of \textcite{bb2000}, and the first-order condition approach of \textcite{gnr2020}.

% Assumptions
To keep the estimation strategies comprehensible and the notation simple, I restrict myself to the case of three inputs ($k_{it}$, $l_{it}$, and $m_{it}$), and make the following assumptions. i) Firms are time-homogeneous in their \textcite{cd1928}-type production technology, i.e. $P(\mathbf{x}_{it}) = (1, k_{it}, l_{it}, m_{it})$ and $\boldsymbol{\beta}_{i} = \boldsymbol{\beta}$. ii) Each firm chooses its period $t$ inputs based on available information $\mathcal{I}_{it}$, where $\omega_{it} \in \mathcal{I}_{it}$ is known before and $\epsilon_{it} \notin \mathcal{I}_{it}$ is realized after each firm's input decisions. iii) Firms predict their future productivity by $\omega_{it} = h_{\omega}^{0}(\omega_{it - 1}) + \eta_{it}$, where $\eta_{it}$ is unknown before $t$. iv) Input choices for capital and labor have dynamic implications, e.g. capital is accumulated by past investment decisions and labor is partially predetermined by hiring and firing costs. 

% Notation
In the following, I denote the $P$-dimensional vector with model parameters as $\boldsymbol{\theta}^{0}$. Furthermore, I refer to $\mathbf{g}(\xi_{i}^{t}, \boldsymbol{\theta}^{0})$ as $P^{\prime}$-dimensional vector of moments, where $\xi_{i}^{t} \coloneqq \{\xi_{it^\prime} \colon t^{\prime} \in \{0, \ldots, t\}\}$, so that $P^{\prime} \geq P$ moment conditions of the form $\EX[\mathbf{g}(\xi_{i}^{t}, \boldsymbol{\theta}^{0})] = 0$ can be used to estimate $\boldsymbol{\theta}^{0}$ by GMM.

% Example 1
\begin{example}
	\textcite{acf2015} propose a control function approach for a specific value-added production function 
	\begin{equation*}
		y_{it} = \beta_{0}^{0} + \beta_{1}^{0} k_{it} + \beta_{2}^{0} l_{it} + \omega_{it} + \epsilon_{it} = \beta_{3}^{0} m_{it} + \omega_{it} + \epsilon_{it} \, .
	\end{equation*}
	This is a reasonable specification if there is perfect complementarity between $m_{it}$ and the other two inputs $k_{it}$ and $l_{it}$, i.e.
	\begin{equation*}
		y_{it} = \min(\beta_{0}^{0} + \beta_{1}^{0} k_{it} + \beta_{2}^{0} l_{it}, \beta_{3}^{0} m_{it}) + \omega_{it} + \epsilon_{it} \, .
	\end{equation*}
	Their identification strategy requires two additional assumptions. First, firms choose $m_{it} = h_{m}^{0}(k_{it}, l_{it}, \omega_{it})$. Second, $h_{m}^{0}(k_{it}, l_{it}, \omega_{it})$ must be a strictly monotonically increasing function in $\omega_{it}$, so that the only unobservable in this function can be expressed as $\omega_{it} = h_{m}^{0\langle-1\rangle}(k_{it}, l_{it}, m_{it})$, where $h_{m}^{0\langle-1\rangle}(\cdot)$ denotes the inverse function of $h_{m}^{0}(\cdot)$. This yields the following system of two equations:
	\begin{align}
		y_{it} =& \; \beta_{0}^{0} + \beta_{1}^{0} k_{it} + \beta_{2}^{0} l_{it} + h_{m}^{0\langle-1\rangle}(k_{it}, l_{it}, m_{it}) + \epsilon_{it} =  h^{0}(k_{it}, l_{it}, m_{it}) + \epsilon_{it} \, , \nonumber \\
		y_{it} =& \; \beta_{0}^{0} + \beta_{1}^{0} k_{it} + \beta_{2}^{0} l_{it} + h_{\omega}^{0}(h^{0}(k_{it - 1}, l_{it - 1}, m_{it - 1}) - \beta_{0}^{0} - \beta_{1}^{0} k_{it - 1} - \beta_{2}^{0} l_{it - 1}) + v_{it} \nonumber \, ,
	\end{align}
	where $h^{0}(\cdot)$ and $h_{\omega}^{0}(\cdot)$ are unknown functions, usually approximated by polynomials, and $v_{it} \coloneqq \eta_{it} + \epsilon_{it}$. The model parameters can then be identified from the following moment conditions: $\EX[\epsilon_{it} \mid \mathcal{I}_{it}] = 0$ and $\EX[v_{it} \mid \mathcal{I}_{it - 1}] = 0$. For instance, if $h^{0}(k_{it}, l_{it}, m_{it}) = \alpha_{0}^{0} + \alpha_{1}^{0} k_{it} + \alpha_{2}^{0} l_{it} + \alpha_{3}^{0} m_{it}$ and $h_{\omega}(\omega_{it - 1}) = \delta^{0} \omega_{it - 1}$ then $\boldsymbol{\theta}^{0} = (\alpha_{0}^{0}, \alpha_{1}^{0}, \alpha_{2}^{0}, \alpha_{3}^{0}, \beta_{0}^{0}, \beta_{1}^{0}, \beta_{2}^{0}, \delta^{0})$ and $\mathbf{g}(\xi_{i}^{t}, \boldsymbol{\theta}^{0}) = (\epsilon_{it}, \epsilon_{it} k_{it}, \epsilon_{it} l_{it}, \epsilon_{it} m_{it}, v_{it}, v_{it} k_{it}, v_{it} k_{it - 1}, v_{it} l_{it - 1}, v_{it} m_{it - 1})$.
\end{example}

% Example 2
\begin{example}
	The dynamic panel estimator of \textcite{bb2000} can be used to estimate the model parameters by imposing $h_{\omega}(\omega_{it - 1}) = \delta^{0} \omega_{it - 1}$. To see this, consider the production function
	\begin{equation*}
		y_{it} = \beta_{0}^{0} + \beta_{1}^{0} k_{it} + \beta_{2}^{0} l_{it} + \beta_{3}^{0} m_{it} + \omega_{it} + \epsilon_{it}
	\end{equation*}
	and its reformulation
	\begin{equation*}
		\Delta_{\delta}^{0}(y_{it}) = (1 - \delta^{0}) \beta_{0}^{0} + \beta_{1}^{0} \Delta_{\delta}^{0}(k_{it}) + \beta_{2}^{0} \Delta_{\delta}^{0}(l_{it}) + \beta_{3}^{0} \Delta_{\delta}^{0}(m_{it}) + w_{it} \, ,
	\end{equation*}
	where $\Delta_{\delta}^{0}(x_{it}) \coloneqq x_{it} - \delta^{0} x_{it - 1}$ and $w_{it} \coloneqq \eta_{it} + \Delta_{\delta}^{0} \epsilon_{it}$. Under the additional assumption that $m_{it}$ has dynamic implications, e.g. due to adjustment costs or due to input market frictions, the model parameters can be identified from the moment conditions $\EX[w_{it} \mid \mathcal{I}_{it - 1}]$. For instance, the identification strategy of \textcite{s2020} implies $\boldsymbol{\theta}^{0} = (\beta_{0}^{0}, \beta_{1}^{0}, \beta_{2}^{0}, \beta_{3}^{0}, \delta^{0})$ and $\mathbf{g}(\xi_{i}^{t}, \boldsymbol{\theta}^{0}) = (w_{it}, w_{it} k_{it}, w_{it} l_{it}, w_{it} k_{it - 1}, w_{it} l_{it - 1}, w_{it} m_{it - 1})$.
\end{example}

% Example 3
\begin{example}
	\textcite{gnr2020} propose an estimation strategy for gross output productions functions, e.g.
	\begin{equation*}
		y_{it} = \beta_{0}^{0} + \beta_{1}^{0} k_{it} + \beta_{2}^{0} l_{it} + \beta_{3}^{0} m_{it} + \omega_{it} + \epsilon_{it} \, .
	\end{equation*}
	In their baseline identification strategy, they assume that firms maximize their expected future profits in a perfectly competitive environment. The idea is to identify $\beta_{3}^{0}$ from firms' optimal intermediate input decisions and then to exploit the dynamic structure of the model to identify the remaining model parameters. Under the additional assumption that $M_{it}$ is a flexible input with a linear cost function $P_{t}^{M} M_{it}$, firms choose $M_{it}$ as the solution to
	\begin{equation*}
		\max_{M_{it} \in \mathbb{R}^{>0}} \EX[P_{t}^{Y} \exp(\beta_{0}^{0} + \beta_{1}^{0} k_{it} + \beta_{2}^{0} l_{it} + \beta_{3}^{0} m_{it} + \omega_{it} + \epsilon_{it}) - P_{t}^{M} M_{it} \mid \mathcal{I}_{it}] \, ,
	\end{equation*} 
	where $P_{t}^{Y}$ and $P_{t}^{M}$ are common output and intermediate input prices, respectively. Reformulating the first-order condition
	\begin{equation*}
		P_{it}^{Y} \exp(\beta_{0}^{0} + \beta_{1}^{0} k_{it} + \beta_{2}^{0} l_{it} + \beta_{3}^{0} m_{it} + \omega_{it}) \beta_{3}^{0} \mathcal{E}^{0} = P_{it}^{M} M_{it}
	\end{equation*}
	and exploiting the dynamic structure yields the following system of two equations:
	\begin{align*}
		s_{it} =& \; \log(\beta_{3}^{0} \mathcal{E}^{0}) - \epsilon_{it} \, , \\
		y_{it}^{r} =& \; \beta_{0}^{0} + \beta_{1}^{0} k_{it} + \beta_{2}^{0} l_{it} + h_{\omega}^{0}(y_{it - 1}^{r} - \beta_{1}^{0} k_{it - 1} - \beta_{2}^{0} l_{it - 1}) + \eta_{it} \, ,
	\end{align*}
	where $s_{it} \coloneqq  \log((P_{it}^{M} M_{it}) / (P_{it}^{Y} Y_{it}))$ is the logarithm of intermediate inputs expenditures relative to revenues, $\mathcal{E}^{0} \coloneqq \EX[\exp(\epsilon_{it}) \mid \mathcal{I}_{it}] = \EX[\exp(\epsilon_{it})]$ is a positive constant, and $y_{it}^{r} \coloneqq y_{it} - \beta_{3}^{0} m_{it} - \epsilon_{it}$. One difference to the other two strategies is that the researcher additionally needs data on the ratio of prices $P_{t}^{M} / P_{t}^{Y}$. The model parameters can then be identified from the following moment conditions: $\EX[\epsilon_{it} \mid \mathcal{I}_{it}] = 0$, $\EX[\exp(\epsilon_{it})] = \mathcal{E}^{0}$, and $\EX[\eta_{it} \mid \mathcal{I}_{it - 1}] = 0$. For instance, if $h_{\omega}(\omega_{it - 1}) = \delta^{0} \omega_{it - 1}$ then $\boldsymbol{\theta}^{0} = (\beta_{3}^{0}, \mathcal{E}^{0}, \beta_{0}^{0}, \beta_{1}^{0}, \beta_{2}^{0}, \delta^{0})$ and $\mathbf{g}(\xi_{i}^{t}, \boldsymbol{\theta}^{0}) = (\epsilon_{it}, \exp(\epsilon_{it}) - \mathcal{E}^{0}, \eta_{it}, \eta_{it} k_{it}, \eta_{it} l_{it}, \eta_{it} y_{it - 1}^{r})$.
\end{example}

\section{Identifying Latent Group Structures}
\label{sec:estimation}

\subsection{Penalized Generalized Method of Moments Estimator}

% Introduction and intuition
I suggest to use the penalized GMM (PGMM) estimator of \textcite{ssp2016} to estimate production functions with $J$ latent groups. The authors propose a novel Lasso penalization technique that achieves classification by shrinking firm-specific towards group-specific model parameters. Given $J$ and a strictly positive tuning parameter $\lambda$, the estimation procedure consists of three subsequent steps. First, the firm- and group-specific model parameters are estimated. Second, based on these estimates, each firm is assigned to a latent group. Third, group-specific model parameters are re-estimated based on the assigned latent group membership. In the following, I explain the three subsequent steps in more detail.

In the first step, the firm- and group-specific model parameters are estimated by minimizing
\begin{equation}
	\label{eq:pgmm}
	Q_{\lambda}^{J}((\boldsymbol{\pi}_{1}, \ldots, \boldsymbol{\pi}_{N}), (\boldsymbol{\theta}_{1}, \ldots, \boldsymbol{\theta}_{J})) \coloneqq \frac{1}{N} \sum_{i = 1}^{N} \overline{\mathbf{g}}(\xi_{i}^{T}, \boldsymbol{\pi}_{i})^{\prime} \, \mathbf{W}_{i} \, \overline{\mathbf{g}}(\xi_{i}^{T}, \boldsymbol{\pi}_{i}) + \lambda \prod_{j = 1}^{J} \norm{\boldsymbol{\pi}_{i} - \boldsymbol{\theta}_{j}} \, ,
\end{equation}
where $\mathbf{W}_{i}$ is a positive definite $P^{\prime} \times P^{\prime}$ weighting matrix, e.g. $\mathbf{W}_{i} = \eye_{P^{\prime}}$, and
\begin{equation}
	\label{eq:moments_firm}
	\overline{\mathbf{g}}(\xi_{i}^{T}, \boldsymbol{\theta}) \coloneqq \frac{1}{T} \sum_{t = 1}^{T} \mathbf{g}(\xi_{i}^{t}, \boldsymbol{\theta}) \, .
\end{equation}
Examples for $\boldsymbol{\pi}_{i}$ and $\mathbf{g}(\xi_{i}^{t}, \boldsymbol{\pi}_{i})$ are provided in Section \ref{sec:firm_moments}, e.g. $\boldsymbol{\pi}_{i} = (\beta_{3i}, \mathcal{E}_{i}, \beta_{0i}, \beta_{1i}, \beta_{2i}, \delta_{i})$ and $\mathbf{g}(\xi_{i}^{t}, \boldsymbol{\pi}_{i}) = (\epsilon_{it}, \exp(\epsilon_{it}) - \mathcal{E}_{i}, \eta_{it}, \eta_{it} k_{it}, \eta_{it} l_{it}, \eta_{it} y_{it - 1}^{r})$ for Example 3. The first term in \eqref{eq:pgmm} is a sum of firm-specific GMM objective functions, and the second term is a penalty term controlled by $\lambda$. The former is minimal when all firm-specific moments are close to zero, whereas the latter is minimal when each $\boldsymbol{\pi}_{i}$ is close to any $\boldsymbol{\theta}_{j}$. Thus, while the GMM part ensures that the model fits the data, the penalty term forces each firm-specific model parameter to be close to any group-specific parameter.

In the second step, each firm is assigned to a latent group using a classification rule proposed by \textcite{ssp2016}:
\begin{equation}
	\label{eq:classification_rule1}
	\widehat{G}_{j} \coloneqq \{i \in \{1, \ldots, N\} \colon \min( \{\norm{\hat{\boldsymbol{\pi}}_{i} - \hat{\boldsymbol{\theta}}_{j^{\prime}}} \colon j^\prime \in \{1, \ldots, J\}) = \norm{\hat{\boldsymbol{\pi}}_{i} - \hat{\boldsymbol{\theta}}_{j}}\} \; \text{for} \; j \in \{1, \ldots, J\} \, ,
\end{equation}
where $(\hat{\boldsymbol{\pi}}_{1}, \ldots, \hat{\boldsymbol{\pi}}_{N})$ and $(\hat{\boldsymbol{\theta}}_{1}, \ldots, \hat{\boldsymbol{\theta}}_{J})$ are the PGMM estimates from the first step. Each firm is assigned to the latent group to which it is most similar, where similarity here is defined as the euclidean distance between the firm- and group-specific PGMM estimates. 

In the third step, the group-specific model parameters are re-estimated by minimizing
\begin{equation}
	\label{eq:post_lasso}
	\widetilde{Q}(\boldsymbol{\theta}_{1}, \ldots, \boldsymbol{\theta}_{J}) \coloneqq \frac{1}{J} \sum_{j = 1}^{J} \Big(\frac{1}{\lvert \widehat{G}_{j} \rvert} \sum_{i \in \widehat{G}_{j}} \overline{\mathbf{g}}(\xi_{i}^{T}, \boldsymbol{\theta}_{j})\Big)^{\prime} \, \mathbf{W}_{j} \, \Big(\frac{1}{\lvert \widehat{G}_{j} \rvert} \sum_{i \in \widehat{G}_{j}} \overline{\mathbf{g}}(\xi_{i}^{T}, \boldsymbol{\theta}_{j})\Big) \, ,
\end{equation}
where $\mathbf{W}_{j}$ is a positive definite $P^{\prime} \times P^{\prime}$ weighting matrix, e.g. $\mathbf{W}_{j} = \eye_{P^{\prime}}$, and $\widehat{G}_{1}, \ldots,  \widehat{G}_{J}$ are the estimated latent groups from the second step. Thus, the third step is simply a standard GMM estimation done separately for each of the estimated latent groups. The corresponding Post-Lasso estimates are denoted as $(\tilde{\boldsymbol{\theta}}_{1}, \ldots, \tilde{\boldsymbol{\theta}}_{J})$.

\begin{remark}
	\textbf{(Group assignment).} The classification rule in the second step ensures that each firm is assigned to a latent group. It is also possible to use a stricter rule that may leave some of the firms unclassified
	\begin{equation}
		\label{eq:classification_rule2}
		\widehat{G}_{j} \coloneqq \{i \in \{1, \ldots, N\} \colon \min( \{\norm{\hat{\boldsymbol{\pi}}_{i} - \hat{\boldsymbol{\theta}}_{j^{\prime}}} \colon j^\prime \in \{1, \ldots, J\}) = \norm{\hat{\boldsymbol{\pi}}_{i} - \hat{\boldsymbol{\theta}}_{j}} \leq \varepsilon\} \; \text{for} \; j \in \{1, \ldots, J\} \, ,
	\end{equation}
	where $\varepsilon$ is small positive constant. A firm is only assigned to its most similar group if the euclidean distance is sufficiently small. Stricter rules can help to deal with outlier firms and thus improve the performance of the estimation procedure. For clarification, unclassified firms are then excluded in the third step.
\end{remark}

\textcite{ssp2016} develop a limiting theory for PGMM estimators that can be applied to dynamic linear panel models such as those that are typically estimated by \textcite{ab1991} estimators. Under asymptotics where $N, T \rightarrow \infty$, but not necessary at the same rate, and under the assumptions that $J = J^{0}$ and $\lambda \in \{T^{- a} \colon a \in (0, 0.5)\}$, the authors show that their PGMM estimator is able to assign all firms belonging to a latent group to the same group with probability approaching one.\footnote{Even if all firms are correctly classified, it does not follow that $\widehat{G}_{j} = G_{j}^{0}$. The classification rule only ensures that $\widehat{G}_{j} \in \{G_{1}^{0}, \ldots, G_{J^{0}}^{0}\}$, i.e. all firms in $G_{j}^{0}$ are assigned to the same latent group.} This classification consistency allows them to show that their Post-Lasso estimator is asymptotically equivalent to an infeasible estimator that knows and exploits the true latent group structure. Due to the close relationship between dynamic panel estimators and the estimation strategies for production functions, as pointed out by \textcite{acf2015}, I conjecture that the properties derived by \textcite{ssp2016} also apply to my proposed nonlinear estimation procedure.\footnote{Although the production function itself is linear in parameters, the assumption about the evolution of productivity causes the entire model to become nonlinear. A formal proof for my estimation procedure will be added later. My conjecture is further supported by simulation experiments in Section \ref{sec:simulations}.}
Thus, given that $(\tilde{\boldsymbol{\theta}}_{1}, \dots, \tilde{\boldsymbol{\theta}}_{J^{0}})$ is a suitable permutation of Post-Lasso estimators, the asymptotic distribution of $\tilde{\boldsymbol{\theta}}_{j}$ can be approximated by $\N(\boldsymbol{\theta}_{j}^{0}, \mathbf{V}_{j})$ for all $j \in \{1, \ldots, J^{0}\}$, where $\mathbf{V}_{j}$ is a $P \times P$ covariance matrix.

\begin{remark}
	\textbf{(GMM inference).} The weighting matrices $\mathbf{W}_{i}$ and $\mathbf{W}_{j}$ affect the performance of my estimation procedure only when the number of moment conditions is larger than the number of model parameters, i.e. $P^{\prime} > P$. Optimal weighting matrices that yield the most efficient GMM estimators are derived in \textcites{h1982}{hs1982}. Since these optimal weighting matrices depend on the model parameters, they have to be estimated. Thus, \textcite{hs1982} suggest a two-step procedure where the optimal weighting matrix is constructed from first-step estimates. For $P^{\prime} > P$, $\mathbf{W}_{i}$ and $\mathbf{W}_{j}$ can also be constructed using two-step procedures, where in the first step $\mathbf{W}_{i} = \mathbf{W}_{j} = \eye_{P^{\prime}}$. Appropriate estimators for $\mathbf{V}_{1}, \ldots, \mathbf{V}_{J^{0}}$ depend on the identification strategy chosen and the underlying structural model assumptions. Examples of common estimators are heteroskedasticity consistent estimators ala \textcite{w1980} or heteroskedasticity and autocorrelation consistent estimators ala \textcite{nw1987}. Alternatively, inference can be based on panel bootstrap procedures like \textcite{k2008}.
\end{remark}

\begin{remark}
	\textbf{(Unbalanced panels).} The estimation procedure can also be applied to unbalanced panels. Assuming that the observations of each firm are consecutive, i.e. $\{\xi_{it} \colon i \in \{1, \ldots, N\}, t \in \{t_{i}, \ldots, T_{i}\}, 0 \leq t_{i} < T_{i} \leq T\}$, adapting the estimator requires only replacing \eqref{eq:moments_firm} with
	\begin{equation}
		\overline{\mathbf{g}}(\xi_{i}^{T_{i}}, \boldsymbol{\theta}) = \frac{1}{T_{i} - t_{i}} \sum_{t = t_{i} + 1}^{T_{i}} \mathbf{g}(\xi_{i}^{t}, \boldsymbol{\theta}) \, .
	\end{equation}
	The asymptotic theory developed by \textcite{ssp2016} applies only to balanced panels, but has recently been extended to unbalanced panels by \textcite{swj2019}. The crucial difference between the theories is that $\min(\{T_{i} - t_{i} \colon i \in {1, \ldots, N}\})$ has to be sufficiently large to ensure classification consistency in unbalanced panels.
\end{remark}

\subsection{Determining the Number of Latent Groups}

The asymptotic theory assumes that the true number of latent groups is known. Since this is very unlikely in practice, I follow \textcite{ssp2016} and suggest to estimate $J^{0}$ by minimizing a BIC-type information criterion
\begin{equation}
	\label{eq:information_criterion}
	\text{IC}_{p}(J, \lambda) \coloneqq \log\Big(\frac{1}{NT}\sum_{j = 1}^{J} \sum_{i \in \widehat{G}_{j}} \sum_{t = 1}^{T} (\tilde{r}_{it}(J, \lambda))^{2}\Big) + J P p(N, T) \, ,
\end{equation}
where $J$ is a guess for the true number of latent groups, $\tilde{r}_{it}(J, \lambda)$ are Post-Lasso residuals of an estimation given $\lambda$ and $J$, e.g. $\tilde{r}_{it}(J, \lambda) = \tilde{\eta}_{it}(J, \lambda) + \tilde{\epsilon}_{it}(J, \lambda)$ for Example 3, and $p(N, T)$ is a penalty term that satisfies $p(N, T) \rightarrow 0$ and $NT p(N, T) \rightarrow \infty$ as $N, T \rightarrow \infty$. \textcite{ssp2016} suggest two penalty terms: $p(N, T) = 2 / 3 \, (NT)^{- 0.5}$ and $p(N, T) = 0.25 \log(\log(T)) / T$. Further examples of suitable penalty terms can be found in \textcites{bn2002}{ln2012}{bm2015}. Thus, given $\lambda$ and $p(N, T)$, the number of latent groups can be estimated as
\begin{equation}
	\label{eq:ic_estimator1}
	\widehat{J}_{p}(\lambda) \in \underset{J \in \{1, \ldots, \overline{J}\} }{\argmin} \text{IC}_{p}(J, \lambda) \, ,
\end{equation}
where $\overline{J}$ is a known upper bound on the true number of latent groups. 

\begin{remark}
	\textbf{(Joint determination).} As pointed out by \textcite{ssp2016}, the information criterion can also be used to jointly determine $\lambda$ and $J$,
	\begin{equation}
		\label{eq:ic_estimator2}
		\widehat{J}_{p} \in \underset{\lambda \in \mathcal{L}}{\argmin} \text{IC}_{p}(\widehat{J}_{p}(\lambda), \lambda) \, ,
	\end{equation}
	where $\mathcal{L} \coloneqq \{T^{- a} \colon a \in (0, 0.5)\}$. In practice, a grid of candidate values between 0 and 0.5 can be used to keep the number of estimates tractable, e.g. $a \in \{0.05, 0.1, 0.15, 0.2, 0.25, 0.3, 0.35, 0.4, 0.45\}$.
\end{remark}

\subsection{Estimation Algorithm}

The PGMM objective function is not jointly convex in $(\boldsymbol{\pi}_{1}, \ldots, \boldsymbol{\pi}_{N})$ and $(\boldsymbol{\theta}_{1}, \ldots, \boldsymbol{\theta}_{J})$ and therefore very costly to minimize. To reduce the computational costs, \textcite{ssp2016} suggest to split the optimization problem into a sequence of $J$ convex subproblems that are solved sequentially until convergence. The objective function of the $j$-th subproblem is 
\begin{equation}
	\label{eq:subproblem}
	Q_{\lambda}^{\langle j \rangle}((\boldsymbol{\pi}_{1}^{\langle j \rangle}, \ldots, \boldsymbol{\pi}_{N}^{\langle j \rangle}), \boldsymbol{\theta}_{j}) \coloneqq \frac{1}{N} \sum_{i = 1}^{N} \overline{\mathbf{g}}(\xi_{i}^{T}, \boldsymbol{\pi}_{i}^{\langle j \rangle})^{\prime} \, \mathbf{W}_{i} \, \overline{\mathbf{g}}(\xi_{i}^{T}, \boldsymbol{\pi}_{i}^{\langle j \rangle}) + \lambda \norm{\boldsymbol{\pi}_{i}^{\langle j \rangle} - \boldsymbol{\theta}_{j}} \zeta_{i}^{\langle j \rangle} \, ,
\end{equation}
where $\zeta_{i}^{\langle j \rangle} \coloneqq \prod_{j^{\prime} \neq j}^{J} \norm{\boldsymbol{\pi}_{i}^{\langle j^{\prime} \rangle} - \boldsymbol{\theta}_{j^{\prime}}}$ is the fixed part of the additive-multiplicative penalty term. The algorithm can be sketched as follows.
\begin{algorithm}
	\textbf{(Iterative Algorithm of \textcite{ssp2016}).}
	\begin{description}
		\item[Step 0.] Given $\lambda$ and $J$, e.g. $\lambda = T^{- 0.25}$ and $J = J^{0}$, initialize $((\boldsymbol{\pi}_{1}^{\langle 1 \rangle}, \ldots, \boldsymbol{\pi}_{N}^{\langle 1 \rangle}), \ldots, (\boldsymbol{\pi}_{1}^{\langle J \rangle}, \ldots, \boldsymbol{\pi}_{N}^{\langle J \rangle}))$ and $(\boldsymbol{\theta}_{1}, \ldots, \boldsymbol{\theta}_{J})$, e.g. $(\boldsymbol{\pi}_{1}^{\langle 1 \rangle}, \ldots, \boldsymbol{\pi}_{N}^{\langle 1 \rangle}) = \ldots = (\boldsymbol{\pi}_{1}^{\langle J \rangle}, \ldots, \boldsymbol{\pi}_{N}^{\langle J \rangle}) = (\boldsymbol{\vartheta}_{1}, \ldots, \boldsymbol{\vartheta}_{N})$ and $(\boldsymbol{\theta}_{1}, \ldots, \boldsymbol{\theta}_{J}) = \mathbf{0}_{P \times J}$, where $(\boldsymbol{\vartheta}_{1}, \ldots, \boldsymbol{\vartheta}_{N})$ are estimates of firm-specific production functions.
		\item[Step 1.] For each $j \in \{1, \ldots, J\}$, compute $\boldsymbol{\zeta}_{i}^{\langle j \rangle}$ given $((\boldsymbol{\pi}_{1}^{\langle 1 \rangle}, \ldots, \boldsymbol{\pi}_{N}^{\langle 1 \rangle}), \ldots, (\boldsymbol{\pi}_{1}^{\langle J \rangle}, \ldots, \boldsymbol{\pi}_{N}^{\langle J \rangle}))$ and $(\boldsymbol{\theta}_{1}, \ldots, \boldsymbol{\theta}_{J})$, and update $(\boldsymbol{\pi}_{1}^{\langle j \rangle}, \ldots, \boldsymbol{\pi}_{N}^{\langle j \rangle})$ and $\boldsymbol{\theta}_{j}$ by solving the $j$-th subproblem.
		\item[Step 2.] Repeat Step 1 until convergence, e.g. stop if $\lvert Q_{\text{outer}}^{\langle r \rangle} - Q_{\text{outer}}^{\langle r - 1 \rangle} \rvert / (Q_{\text{outer}}^{\langle r - 1 \rangle} + 1) < \varepsilon$, where $Q_{\text{outer}}^{\langle r \rangle} \coloneqq \sum_{j = 1}^{J} Q_{\lambda}^{\langle j \rangle}((\boldsymbol{\pi}_{1}^{\langle j, \, r \rangle}, \ldots, \boldsymbol{\pi}_{N}^{\langle j, \, r \rangle}), \boldsymbol{\theta}_{j}^{\langle r \rangle})$ is the sum of the function values of all subproblems in the $r$-th iteration and $\varepsilon$ is a small positive constant.
	\end{description}
\end{algorithm}

Although convexity significantly reduces the computational cost, the optimization problem is still challenging because the solution of each subproblem in Step 1 involves $(N + 1) \times P$ parameters. For instance, in my empirical illustration, I use a panel of $N = 571$ firms and a estimation strategy with $P = 6$ model parameters, implying a total of $3{,}432$ parameters per subproblem. To further reduce the computational costs, I suggest an \textit{Alternate Convex Search} (see \cite{htw2015}, chapter 5.9) that exploits the separable structure of \eqref{eq:subproblem}. Instead of minimizing jointly over $(\boldsymbol{\pi}_{1}^{\langle j \rangle}, \ldots, \boldsymbol{\pi}_{N}^{\langle j \rangle})$ and $\boldsymbol{\theta}_{j}$, I can alternate between solving two optimization problems: i) minimization over $(\boldsymbol{\pi}_{1}^{\langle j \rangle}, \ldots, \boldsymbol{\pi}_{N}^{\langle j \rangle})$ holding $\boldsymbol{\theta}_{j}$ fixed and ii) minimization over $\boldsymbol{\theta}_{j}$ holding $(\boldsymbol{\pi}_{1}^{\langle j \rangle}, \ldots, \boldsymbol{\pi}_{N}^{\langle j \rangle})$ fixed. Minimization i) is still over $N \times P$ parameters. However, holding $\boldsymbol{\theta}_{j}$ fixed, the optimization problem can be separated into $N$ independent subproblems involving only $P$ parameters each. Minimization ii) involves only $P$ parameters and is just a Fermat-Weber location problem for which there are numerous efficient algorithms, e.g. the fixed-point algorithm of \textcite{w1937}. The algorithm can be sketched as follows.
\begin{algorithm}
	\textbf{(Alternate Convex Search).}
	\begin{description}
		\item[Step 0.] Given $\lambda$, $j$, and $(\zeta_{1}^{\langle j \rangle}, \ldots, \zeta_{N}^{\langle j \rangle})$, initialize $(\boldsymbol{\pi}_{1}^{\langle j \rangle}, \ldots, \boldsymbol{\pi}_{N}^{\langle j \rangle})$ and $\boldsymbol{\theta}_{j}$.
		\item[Step 1.] Given $(\boldsymbol{\pi}_{1}^{\langle j \rangle}, \ldots, \boldsymbol{\pi}_{N}^{\langle j \rangle})$ update $\boldsymbol{\theta}_{j}$ as
		\begin{equation*}
			\underset{\boldsymbol{\theta} \in \mathbb{R}^{P}}{\argmin} \frac{\lambda}{N} \sum_{i = 1}^{N} \norm{\boldsymbol{\pi}_{i}^{\langle j \rangle} - \boldsymbol{\theta}} \zeta_{i}^{\langle j \rangle} \, .
		\end{equation*}
		\item[Step 2.] Given $\boldsymbol{\theta}_{j}$, for each $i \in \{1, \ldots, N\}$,  update $\boldsymbol{\pi}_{i}^{j}$ as
		\begin{equation*}
			\underset{\boldsymbol{\pi} \in \mathbb{R}^{P}}{\argmin} \overline{\mathbf{g}}(\xi_{i}^{T}, \boldsymbol{\pi})^{\prime} \, \mathbf{W}_{i} \, \overline{\mathbf{g}}(\xi_{i}^{T}, \boldsymbol{\pi}) + \lambda \norm{\boldsymbol{\pi} - \boldsymbol{\theta}_{j}} \zeta_{i}^{\langle j \rangle} \, .
		\end{equation*}
		\item[Step 3.] Repeat Step 1 and Step 2 until convergence, e.g. stop if $\lvert Q_{\text{inner}}^{\langle r \rangle} - Q_{\text{inner}}^{\langle r - 1 \rangle} \rvert / (Q_{\text{inner}}^{\langle r - 1 \rangle} + 1) < \varepsilon$, where $Q_{\text{inner}}^{\langle r \rangle} \coloneqq Q_{\lambda}^{\langle j \rangle}((\boldsymbol{\pi}_{1}^{\langle j, \, r \rangle}, \ldots, \boldsymbol{\pi}_{N}^{\langle j, \, r \rangle}), \boldsymbol{\theta}_{j}^{\langle r \rangle})$ is the function value of the $j$-th subproblem in the $r$-th iteration and $\varepsilon$ is a small positive constant.
	\end{description}
\end{algorithm}

\begin{remark}
	\textbf{(Computation).}	First, although each minimization problem in Step 2 is convex, the corresponding objective function is not differentiable due to the Euclidean norm in the penalty function. For these types of minimization problems, there are special optimization algorithms, such as those presented in \textcite[][chapter 5]{htw2015}. Second, because the $N$ subproblems are independent of each other, they can also be easily parallelized.
\end{remark}

\section{Simulation Experiments}
\label{sec:simulations}

To analyze the classification accuracy and statistical properties of my proposed estimation procedure, I extend the representative firm model of \textcite{s2001} to incorporate latent group structures. The model has the advantage that the input decision problem of each firm can be solved analytically. Similar firm models have been used, for instance, by \textcites{vb2007}{acf2015}{cwdl2016}.

I simulate samples of $N = 200$ firms. Each firm is observed for $T$ time periods and belongs to one of $J^{0} = 3$ latent groups. Each group consists of $N_{j}$ firms. All firms are time-homogeneous within a group, but differ in their parameter configuration and in their relative occurrence between groups. At the beginning of period $t$, a firm $i$ with rational expectations has the following input decision problem:
\begin{eqnarray}
	&&\max_{M_{it},\, I_{it}} \, \EX \Big[\sum_{t = 0}^{\infty} b^{t} (Y_{it} - M_{it} - \phi_{i} I_{it}^{2} / 2) \mid \mathcal{I}_{it}\Big]  \\
	&&\text{subject to} \nonumber \\
	&&Y_{it} \coloneqq K_{it}^{\beta_{i}} M_{it}^{\gamma_{i}} \exp(\omega_{it} + \epsilon_{it}) \, , \nonumber \\
	&&K_{it} \coloneqq (1 - d) K_{it - 1} + I_{it - 1} \, , \nonumber \\
	&&\omega_{it} \coloneqq \alpha_{i} + \delta_{i} \omega_{it - 1} + \eta_{it} \, , \nonumber
\end{eqnarray}
where $Y_{it}$, $K_{it}$, $M_{it}$, $I_{it}$ are output, capital, intermediate input, and investment, $\mathcal{I}_{it}$ is a set of information available at the beginning of period $t$, $\omega_{it}$ is an anticipated productivity shock, and $\epsilon_{it}$ is an unanticipated productivity shock realized after each firm's input decision. Furthermore, $\log(\phi_{i}^{- 1}) \sim \iid \N(0, 1)$, $\eta_{it} \sim \iid \N(0, \sigma_{\eta_{i}}^{2})$, $\epsilon_{it} \sim \iid \N(0, \sigma_{\epsilon_{i}}^{2})$, $\omega_{i0} \sim \iid \N(\alpha_{i} / (1 - \delta_{i}), \sigma_{\eta_{i}}^{2} / (1 - \delta_{i}^2))$, and $K_{i0} \coloneqq \mathbf{0}_{N}$. The model parameters are described and defined in Table \ref{tab:configuration}.
\begin{table}[!htbp]
	\centering
	\begin{threeparttable}
		\caption{Model Parameters: Description and Definition}
		\label{tab:configuration}
		\begin{tabular}{@{}*{2}{l}*{3}{c}@{}}
			\toprule
			Parameter&Description&\multicolumn{3}{c}{Latent Group}\\
			\cmidrule(lr){3-5}
			&&1&2&3\\
			\midrule
			\multicolumn{5}{l}{\textbf{Heterogeneous Parameters:}}\\[0.5em]
			$\quad N_{j} / N$&Relative Occurrence&0.30&0.40&0.30\\
			$\quad \gamma_{i}$&Output Elasticity of Intermediate Input&0.35&0.50&0.65\\
			$\quad \beta_{i} \coloneqq 1 -  \gamma_{i}$&Output Elasticity of Capital&0.65&0.50&0.35\\
			$\quad \sigma_{\epsilon_{i}}$&Standard Deviation of Ex-Post Productivity Shock&0.02&0.04&0.02\\
			$\quad \alpha_{i}$&Constant of AR(1) Process&0.00&0.20&0.40\\
			$\quad \delta_{i}$&Slope Parameter of AR(1) Process&0.90&0.80&0.70\\
			$\quad \sigma_{\eta_{i}}$&Standard Deviation of Innovation in AR(1) Process&0.01&0.01&0.01\\
			\multicolumn{5}{l}{\textbf{Homogeneous Parameters:}}\\[0.5em]
			$\quad b$&Discount Factor&\multicolumn{3}{c}{0.985}\\
			$\quad d$&Depreciation Rate&\multicolumn{3}{c}{0.100}\\
			\bottomrule
		\end{tabular}
	\end{threeparttable}
\end{table}  

I briefly summarize the core features of the model. Firms maximize their expected future profits with respect to their intermediate input and investment choices. Future profits are discounted and all firms face firm-specific but time constant quadratic capital adjustment costs, as in \textcite{acf2015}. Within a latent group, all firms share the same time-homogeneous Cobb-Douglas production function with constant returns to scale. The current capital stock is accumulated through a dynamic process determined by depreciation and past investment decisions. Productivity is additively separable and can be decomposed into a persistent component and an ex-post shock. The persistent component can be predicted by an AR(1) process. 

Because $M_{it}$ is a fully flexible input, i.e. there are no adjustment costs or other dynamic implications, firm $i$'s optimal choice at time $t$ follows immediately from the first-order condition, i.e.
\begin{equation*}
	M_{it}^{\ast} \coloneqq (\gamma_{i} \exp(\omega_{it}) \mathcal{E}_{i})^{\frac{1}{\beta_{i}}} \, K_{it} \, ,
\end{equation*}
where $\mathcal{E}_{i} = \exp(\sigma_{\epsilon_{i}}^{2} / 2)$. Using the Euler equation for investment along with forward substitution yields a fully deterministic optimal investment decision 
\begin{equation*}
	I_{it}^{\ast} \coloneqq b \beta_{i} (\gamma_{i} \mathcal{E}_{i})^{\frac{\gamma_{i}}{\beta_{i}}} \phi_{i}^{- 1}
	\sum_{\tau = 0}^{\infty} \Big((b (1 - d))^{\tau} \exp\Big(\frac{\alpha_{i} \sum_{s = 0}^{\tau} \delta_{i}^{s} + \delta_{i}^{\tau + 1}\omega_{it}}{\beta_{i}} + \frac{\sigma_{\eta_{i}}^{2} \sum_{s = 0}^{\tau} \delta_{i}^{2 s}}{2 \beta_{i}^{2}}\Big)\Big) \, .
\end{equation*}
To ensure that all firms are sampled from their steady state distribution, I extend each time span by $1{,}000$ initial periods that are excluded from the final sample. For clarification, the final sample has $N(T + 1)$ observations. The additional time period per firm is used to generate lagged values of the output and input variables so that the final sample used for the estimation has $NT$ observations. Because the optimal investment decision is a convergent series, I approximate it by the first $1{,}001$ terms of the sum.\footnote{More specifically, I split the series into two parts: 
\begin{equation*}
	I_{it}^{\ast} = c_{i} \Big(\sum_{\tau = 0}^{1000} a_{i\tau} + \sum_{\tau = 1001}^{\infty} a_{i\tau}\Big)\, ,
\end{equation*}
where $c_{i}$ is the factor in front of the series and $a_{i\tau}$ are the terms of the series. I assume that the first $1{,}001$ terms are sufficient to approximate the optimal level of investment, i.e. all remaining terms are negligible small.} 

The length of the panel is the key determinant for the performance of my estimation procedure. Thus, I focus on samples with different time spans $T \in \{15, 25, 50\}$. The estimation procedure is based on the identification strategy of \textcite{gnr2020}. The corresponding moment conditions are adjusted to the functional forms of the data generating process, i.e. Cobb-Douglas production function, AR(1) process for the persistent component of productivity, and $s_{it} = m_{it} - y_{it}$. I do not exploit the constant returns to scale restriction that would allow me to recover both elasticities from the share equation only. The analysis consists of two parts. In the first part, I assume that $J^{0}$ is unknown and study the accuracy of different estimators for it. In the second part, I assume that $J = J^{0} = 3$ is known and analyze the finite sample behavior of my estimation procedure. All reported results are based on 100 simulated samples and $\lambda = T^{- 0.25}$.\footnote{In a preliminary analysis, I tried different values for $\lambda \in \{T^{- a} \colon a \in \{0.05, 0.1, 0.15, 0.2, 0.25, 0.3, 0.35, 0.4, 0.45\}\}$ and found that the estimation procedure is very robust to different choices of $\lambda$.}

In the first part, I analyze the performance of $\widehat{J}_{p}(\lambda)$ defined in \eqref{eq:ic_estimator1} with $\overline{J} = 5$. I consider different penalty terms: $p_{1, r}(N, T) \coloneqq r \, (NT)^{- 0.5}$ and $p_{2, r}(N, T) \coloneqq r \log(\log(T)) / T$, where $r \in \{0.25, 0.5, 0.75, 1\}$ is a finite sample adjustment factor.\footnote{I also tried some of the penalty terms suggested by \textcite{bn2002}, but I did not find any better alternatives.} The analysis is based on the following quantities: expected value and probabilities to select exactly or at least $J^{0} = 3$ latent groups. Table \ref{tab:number_of_groups} summarizes the results.
\begin{table}[!htbp]
	\centering
	\begin{threeparttable}
		\caption{Point Estimation of $\widehat{J}_{p}(\lambda)$}
		\label{tab:number_of_groups}
		\begin{tabular}{@{}*{2}{l}*{8}{c}@{}}
			\toprule
			$T$&Quantity&\multicolumn{4}{c}{$p_{1, r}(N, T)$ with $r=$}&\multicolumn{4}{c}{$p_{2, r}(N, T)$ with $r=$}\\
			\cmidrule(lr){3-6}\cmidrule(lr){7-10}
			&&0.25&0.5&0.75&1&0.25&0.5&0.75&1\\
			\midrule
			15&$\EX[\widehat{J}_{p}(\lambda)]$& 3.05 & 3.00 & 3.00 & 3.00 & 3.00 & 3.00 & 2.94 & 2.04 \\
			&$\Pr(\widehat{J}_{p}(\lambda) = 3)$&0.95 & 1.00 & 1.00 & 1.00 & 1.00 & 1.00 & 0.94 & 0.04 \\ 
			&$\Pr(\widehat{J}_{p}(\lambda) \geq 3)$& 1.00 & 1.00 & 1.00 & 1.00 & 1.00 & 1.00 & 0.94 & 0.04\\ 
			\midrule
			25&$\EX[\widehat{J}_{p}(\lambda)]$& 3.00 & 3.00 & 3.00 & 3.00 & 3.00 & 3.00 & 3.00 & 3.00 \\ 
			&$\Pr(\widehat{J}_{p}(\lambda) = 3)$& 1.00 & 1.00 & 1.00 & 1.00 & 1.00 & 1.00 & 1.00 & 1.00 \\ 
			&$\Pr(\widehat{J}_{p}(\lambda) \geq 3)$& 1.00 & 1.00 & 1.00& 1.00 & 1.00 & 1.00 & 1.00 & 1.00 \\ 
			\midrule
			50&$\EX[\widehat{J}_{p}(\lambda)]$& 3.00 & 3.00 & 3.00 & 3.00 & 3.00 & 3.00 & 3.00 & 3.00 \\ 
			&$\Pr(\widehat{J}_{p}(\lambda) = 3)$& 1.00 & 1.00 & 1.00 & 1.00 & 1.00 & 1.00 & 1.00 & 1.00 \\ 
			&$\Pr(\widehat{J}_{p}(\lambda) \geq 3)$&1.00 & 1.00 & 1.00 & 1.00 & 1.00 & 1.00 & 1.00 & 1.00 \\ 
			\bottomrule
		\end{tabular}
		\begin{tablenotes}
			\footnotesize
			\item \emph{Note:} $N = 200$, $J^{0} = 3$, and $\lambda = T^{- 0.25}$; $\widehat{J}_{p}(\lambda)$ is defined in \eqref{eq:ic_estimator1} with $\overline{J} = 5$, $p_{1, r}(N, T) = r \, (NT)^{- 0.5}$, and $p_{2, r}(N, T) = r \log(\log(T)) / T$; results are based on 100 simulated samples.
		\end{tablenotes}
	\end{threeparttable}
\end{table}
For $T \in \{25, 50\}$, all estimators perfectly predict $J^{0}$. Only for $T = 15$, I find noticeable differences in the performance. While the estimator with $p_{1, r}(N, T)$ slightly overestimates the number of latent groups for small $r$, the estimator with $p_{2, r}(N, T)$ underestimates it for large $r$. Given a suitable adjustment factor, estimators based on both specifications are able to perfectly predict $J^{0}$. Overall, my simulation experiments suggest a larger adjustment factor for $p_{1, r}(N, T)$ and a smaller factor for $p_{2, r}(N, T)$, e.g. $r = 1$ for the former and $r = 0.25$ for the latter.

In the second part, I take $J = J^{0}$ as given and compare the finite sample performance of my Post-Lasso estimator (Post-Lasso) with an infeasible estimator that knows and exploits the true latent group structure (Infeasible). For the comparison, I consider the following quantities: bias, standard deviation, root mean squared error, ratio of standard error and standard deviation, and coverage rates of confidence intervals with 95\% nominal level. The statistics are computed separately for each firm. The bias, standard deviation, and root mean squared error are all relative to the true parameter value in percent. To keep the analysis concise, I focus on the output elasticities, i.e. $\boldsymbol{\gamma} \coloneqq (\gamma_{1}, \ldots, \gamma_{N})$ and $\boldsymbol{\beta} \coloneqq (\beta_{1}, \ldots, \beta_{N})$, and report aggregate rather than firm-specific quantities, e.g. relative bias as $100 / N \sum_{i = 1}^{N} (\bar{\hat{\gamma}}_{i} - \gamma_{i}) / \gamma_{i}$, where $\bar{\hat{\gamma}}_{j}$ is the average estimate of $\gamma_{i}$ over all simulated samples. Since the performance of Post-Lasso depends on the classification accuracy in the first step of the estimation procedure, I additionally report the fraction of correctly classified firms. Table \ref{tab:properties} summarizes the results.
\begin{table}[!htbp]
	\centering
	\begin{threeparttable}
		\caption{Classification Accuracy and Point Estimation of $\boldsymbol{\beta}$ and $\boldsymbol{\gamma}$}
		\label{tab:properties}
		\begin{tabular}{@{}*{2}{l}*{4}{c}@{}}
			\toprule
			$T$&Quantity&\multicolumn{2}{c}{Post-Lasso}&\multicolumn{2}{c}{Infeasible}\\
			\cmidrule(lr){3-4}\cmidrule(lr){5-6}
			&&$\boldsymbol{\gamma}$&$\boldsymbol{\beta}$&$\boldsymbol{\gamma}$&$\boldsymbol{\beta}$\\
			\midrule
			15& \% of Correct Classification&\multicolumn{2}{c}{0.9959}&\multicolumn{2}{c}{-}\\
			&Bias (relative in \%)&-0.0476&0.8605&-0.0439&-0.0491\\
			&Standard Deviation (relative in \%) &0.5304&4.3252&0.5117&3.6788\\
			&Root Mean Squared Error (relative in \%)&0.5301&4.4658&0.5136&3.6775\\
			&Standard Error / Standard Deviation&0.9891&0.9513&1.0184&1.0112\\
			&Coverage Rate (nominal level = 0.95)&0.9450&0.9340&0.9560&0.9410\\
			\midrule
			25& \% of Correct Classification&\multicolumn{2}{c}{0.9995}&\multicolumn{2}{c}{-}\\
			&Bias (relative in \%)&0.0194&0.1695&0.0222&0.0216\\
			&Standard Deviation (relative in \%) &0.4087&2.8645&0.4053&2.7452\\
			&Root Mean Squared Error (relative in \%)&0.4104&2.8554&0.4068&2.7367\\
			&Standard Error / Standard Deviation&0.9950&1.0349&1.0033&1.0662\\
			&Coverage Rate (nominal level = 0.95)&0.9690&0.9560&0.9720&0.9590\\
			\midrule
			50& \% of Correct Classification&\multicolumn{2}{c}{1.0000}&\multicolumn{2}{c}{-}\\
			&Bias (relative in \%)&-0.0083&-0.0260&-0.0083&-0.0260\\
			&Standard Deviation (relative in \%) &0.2696&2.0006&0.2696&2.0006\\
			&Root Mean Squared Error (relative in \%)&0.2689&2.0048&0.2689&2.0048\\
			&Standard Error / Standard Deviation&1.0509&1.0339&1.0509&1.0339\\
			&Coverage Rate (nominal level = 0.95)&0.9570&0.9500&0.9570&0.9500\\
			\bottomrule
		\end{tabular}
		\begin{tablenotes}
			\footnotesize
			\item \emph{Note:} $N = 200$, $J = 3$, and $\lambda = T^{- 0.25}$; Post-Lasso is defined in \eqref{eq:post_lasso}; Infeasible knows and exploits the true latent group structure; results are based on 100 simulated samples.
		\end{tablenotes}
	\end{threeparttable}
\end{table}
I start with the first step classification accuracy. The fractions of correctly classified firms are always close to one, even for $T = 15$. The classification accuracy improves as $T$ increases. For $T = 50$, the fraction of correctly classified firms is one. The excellent classification accuracy is also reflected in the finite sample performance. Even for $T = 15$, the performance of Post-Lasso is nearly identical to that of the benchmark estimator. In general, I find biases smaller than 1\%, ratios of standard errors and standard deviation near one, and coverage rates close to their nominal values. Misclassification in the first step mainly results in larger dispersions. Consequently, the largest differences in performance are apparent in the standard deviation and the root mean squared error.

\section{Empirical Illustration}
\label{sec:illustration}

It is standard practice to estimate separate production functions for each industry. In most studies, industries are delineated by an industry classification, like ISIC, SIC, NACE, or NAICS.\footnote{Some recent and influential examples are \textcites{adkpvr2020}{dleu2020}. The former use a 2-digit SIC and the latter use a 2- and 4-digit NAICS classification to define an industry.} The underlying assumption here is that firms operating in the same economic environment, like an industry, use the same production technology. Although ex-ante classification by industry is very convenient and intuitive, the question arises whether this classification is sufficient to fully account for latent firm heterogeneity.

I use a balanced subsample of the Chilean panel data set used by \textcite{gnr2020}.\footnote{Previous studies that also use the Chilean data set include \textcites{p2002}{lp2003}.} The data is provided by \textit{Instituto Nacional de Estadística de Chile} and includes all manufacturing plants with more than ten employees in the five largest industrial sectors, i.e. food products (311), textiles (321), apparel (322), wood products (331), and fabricated metals (381), that were in operation between 1979 and 1996.\footnote{The data is taken from the \href{https://www.journals.uchicago.edu/doi/suppl/10.1086/707736}{replication package} provided by the authors. \href{https://unstats.un.org/unsd/classifications/Econ/ISIC}{\textit{ISIC Rev. 2}} industry classification.} In addition to the output and input variables $(Y_{it}, K_{it}, L_{it}, M_{it}, S_{it})$, the sample contains further firm characteristics. I know whether a firm is an exporter, is an importer of intermediate goods, pays above median industry wages, or is an advertiser. Further, $Y_{it}$ is measured as deflated revenue, $K_{it}$ is the capital stock constructed by the perpetual inventory method, $L_{it}$ is a weighted sum of unskilled and skilled number of workers, $M_{it}$ is measured as the sum of several intermediate input expenditures, e.g. raw materials, energy, and services, and $S_{it}$ is intermediate input expenditure relative to revenue.\footnote{Further details about the construction of the sample are provided in \textcite[][footnote 42]{gnr2020}.} 

I consider the following time-homogeneous Cobb-Douglas production function
\begin{equation}
	\label{eq:prodfun_illustration}
	y_{it} = \beta_{0i} + \beta_{1i} k_{it} + \beta_{2i} l_{it}  + \beta_{3i} m_{it} + \omega_{it} + \epsilon_{it} \, ,
\end{equation}
where $i \in \{1, \ldots, 571\}$, $t \in \{0, \ldots, 17\}$, $\omega_{it} = \delta_{0i} + \delta_{1i} \omega_{it - 1} + \eta_{it}$ is a persistent productivity shock following an AR(1) process, $\eta_{it}$ is an unexpected innovation, and $\epsilon_{it}$ is an ex-post productivity shock. I assume that the model parameters follow an unknown general group pattern. To deal with this latent firm heterogeneity, I consider two classification strategies: i) ex-ante classification by industry and ii) data-driven classification using PGMM estimation. Estimates of both strategies are based on the moment conditions of \textcite{gnr2020}, i.e. the moment conditions defined in Example 3.

\begin{figure}[!htbp]
	\centering
	\caption{Information Criteria 1 and 2 for Different Values of $\lambda$ and $J$}
	\includegraphics{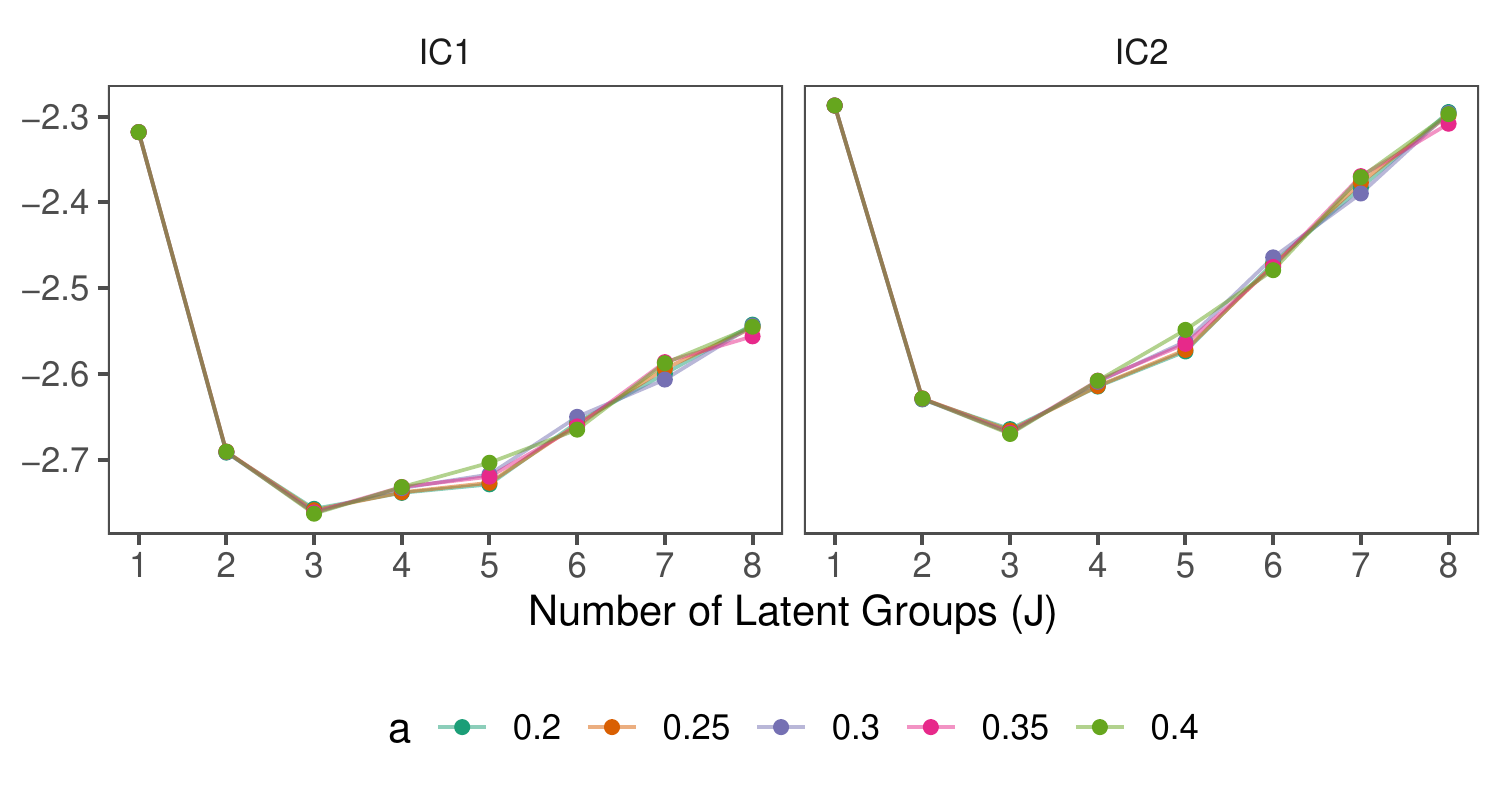}
	\begin{minipage}{\textwidth}
		\footnotesize
		\emph{Note:} $N = 571$, $T = 17$, $\lambda = T^{- a}$, and $J \in \{1, \ldots, 8\}$; IC1 and IC2 are computed based on \eqref{eq:information_criterion} with $p(N, T) = (NT)^{- 0.5} \approx 0.0101$ and $p(N, T) = 0.25 \log(\log(T)) / T \approx 0.0153$, respectively.\\
		\emph{Source:} Data taken from the replication package of \textcite{gnr2020}.
	\end{minipage}
	\label{fig:information_criteria}
\end{figure}

For the PGMM estimation, I need to specify two additional parameters: $\lambda$ and $J$. To determine both parameters jointly, I use information criteria with two different penalty terms: $p(N, T) = (NT)^{- 0.5} \approx 0.0101$ and $p(N, T) = 0.25 \log(\log(T)) / T \approx 0.0153$. I compute both criteria for all combinations of $\lambda \in \{T^{- a} \colon a \in \{0.2, 0.25, 0.3, 0.35, 0.4\}\}$ and $J \in \{1, \ldots, 8\}$, where $J = 1$ refers to the absence of latent firm heterogeneity, i.e. firm homogeneity. Figure \ref{fig:information_criteria} visualizes the results.
Both information criteria suggest $\lambda = 17^{-0.4} \approx 0.3220$ and $J = 3$ for the PGMM estimation. Thus, there is strong evidence for latent firm heterogeneity, as indicated by the significantly improved fit of the structural model for $J > 1$.

\begin{figure}[!htbp]
	\centering
	\caption{Relationship between Ex-Ante and Data-Driven Classification}
	\includegraphics{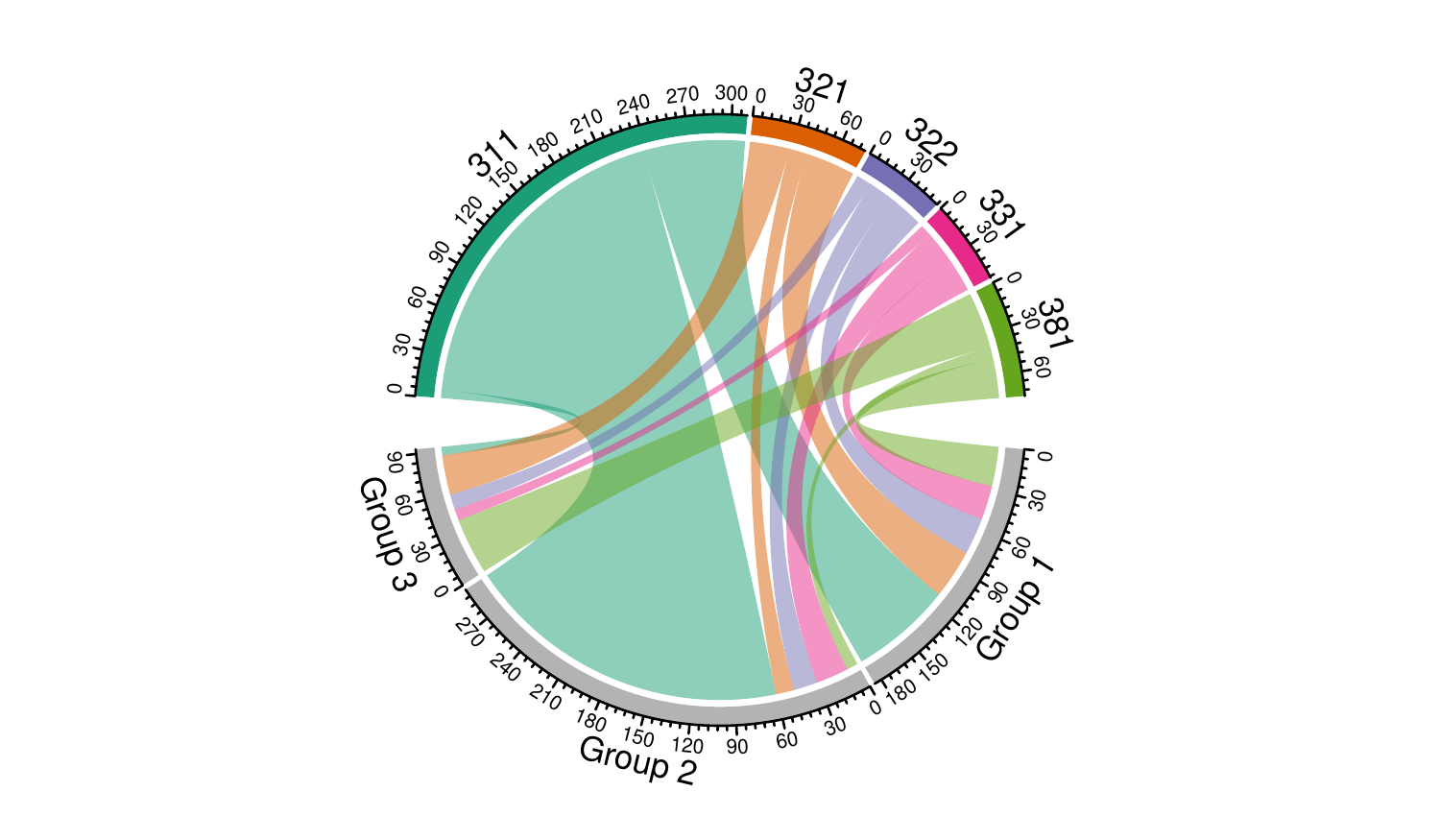}
	\begin{minipage}{\textwidth}
		\footnotesize
		\emph{Note:} $N = 571$, $T = 17$, $\lambda = 17^{-0.4} \approx 0.3220$, and $J = 3$; 311, 321, 322, 331, and 381 are industries; Group 1--3 are the estimated latent groups; the smaller numbers between 0 and 300 indicate firms in each group; the links in different colors represent the connection between an industry sector and a latent group; colors are based on industry sectors.\\
		\emph{Source:} Data taken from the replication package of \textcite{gnr2020}.
	\end{minipage}
	\label{fig:classification}
\end{figure}

So far, I found that the estimated number of latent groups is smaller than the number of industries. However, it is unclear to which extent the data-driven classification matches the ex-ante classification. The chord diagram in Figure \ref{fig:classification} visualizes the matching.
First, I find firms from all three latent groups in each of the five industries. Second, I find that firms from different industries split very unequally among the latent groups. For instance, firms in industries 311 and 381 split mainly between two latent groups, whereas firms in the other industries split more evenly between all three latent groups. Furthermore, about three quarters of the firms in industry 311 are assigned to Group 2. Thus, the matching analysis provides evidence that an ex-ante classification by industry is not sufficient to fully account for latent firm heterogeneity, which is in line with \textcite{kss2017} who report substantial firm heterogeneity even in narrowly defined industries.

Finally, I analyze the extent to which estimates based on ex-ante and data-driven classification differ. I start with the model fit and compare the mean squared residuals (MSR) for both classification approaches. The residuals are defined as $\hat{\eta}_{it} + \hat{\epsilon}_{it}$. The MSR for the industry classification ($\approx 0.0703$) is larger than the MSR for the data-driven classification ($\approx 0.0528$). Thus, the data-driven classification yields a better model fit and does so even with fewer parameters. Next, I analyze the differences in the estimated output elasticities and how these differences translate into heterogeneity in total factor productivity (TFP).\footnote{In addition to output elasticities, TFP is another quantity of interest in some empirical studies. These studies are particularly interested in getting a better understanding of persistent TFP differences that are frequently reported (see \cite{bd2000} and \cite{s2011} for comprehensive overviews).} I follow \textcite{op1996} and estimate TFP in levels as  $\exp(y_{it} - \hat{\beta}_{1i} k_{it} - \hat{\beta}_{2i} l_{it} - \hat{\beta}_{3i} m_{it})$. I analyze the heterogeneity in TFP through excluded firm characteristics using a pseudo-poisson estimator with the following conditional mean specification
\begin{equation}
	\label{eq:tfp_illustration}
	\exp(\pi_{1i} \, \text{trade}_{it} + \pi_{2i} \, \text{highwage}_{it} + \pi_{3i} \, \text{advertiser}_{it} + \alpha_{i} + \gamma_{t}) \, ,
\end{equation}
where the parameters $\pi_{1i}$, $\pi_{2i}$, and $\pi_{3i}$ follow the group pattern implied by the ex-ante or data-driven classification, $\text{trade}_{it} \coloneqq \max(\text{exporter}_{it}, \text{importer}_{it})$, $\text{highwage}_{it}$, and $\text{advertiser}_{it}$ are indicator variables equal to one, if firm $i$ at time $t$ engages in international trade, either through exporting and/or importing, pays above median wages, and is an advertiser, respectively, and $\alpha_{i}$ and $\gamma_{t}$ are firm and year fixed effects. Table \ref{tab:estimates} summarizes the results.  
\begin{sidewaystable}[!htbp]
	\centering
	\begin{threeparttable}
		\caption{Estimation Results: Output Elasticities and Heterogeneity in TFP}
		\label{tab:estimates}
		\begin{tabular}{@{}*{1}{l}*{8}{c}@{}}
			\toprule
			&\multicolumn{5}{c}{Ex-Ante Classification}&\multicolumn{3}{c}{Data-Driven Classification}\\
			\cmidrule(lr){2-6}\cmidrule(lr){7-9}
			&311&321&322&331&381&Group 1&Group 2&Group 3\\
			\midrule
			\multicolumn{9}{l}{\textbf{Output Elasticities:}}\\[0.5em]
			$\quad$Capital  & 0.163 & 0.096 & 0.149 & 0.095 & 0.186 & 0.158 & 0.137 & 0.199 \\ 
			& (0.012) & (0.027) & (0.033) & (0.025) & (0.034) & (0.014) & (0.010) & (0.023) \\ 
			$\quad$Labor & 0.182 & 0.346 & 0.249 & 0.284 & 0.472 & 0.299 & 0.155 & 0.503 \\ 
			& (0.018) & (0.037) & (0.046) & (0.037) & (0.056) & (0.022) & (0.015) & (0.041) \\ 
			$\quad$Intermediates & 0.674 & 0.494 & 0.550 & 0.583 & 0.439 & 0.557 & 0.720 & 0.386 \\ 
			& (0.003) & (0.007) & (0.007) & (0.007) & (0.006) & (0.003) & (0.002) & (0.005) \\ 
			\multicolumn{9}{l}{\textbf{\textbf{Heterogeneity in Total Factor Productivity:}}}\\[0.5em]
			$\quad$Trade & 0.027 & 0.047 & 0.043 & -0.036 & 0.028 & 0.042 & 0.023 & 0.070\\
			& (0.016) & (0.026) & (0.032) & (0.039) & (0.025) & (0.024) & (0.011) & (0.036)\\
			$\quad$Wages > median & 0.042 & 0.053 & 0.085 & 0.070 & 0.070 & 0.043 & 0.048 & 0.076\\
			& (0.008) & (0.024) & (0.033) & (0.033) & (0.027) & (0.016) & (0.009) & (0.039)\\
			$\quad$Advertiser & -0.008 & 0.008 & -0.060 & -0.003 & -0.027 & -0.031 & -0.003 & -0.022\\
			& (0.009) & (0.023) & (0.035) & (0.019) & (0.022) & (0.015) & (0.008) & (0.028)\\
			\bottomrule
		\end{tabular}
		\begin{tablenotes}
			\footnotesize
			\item \emph{Note:} Output Elasticities are estimated based on the moment conditions of \textcite{gnr2020}; \textcite{w1980}-type standard errors in parentheses; Heterogeneity in TFP is analyzed through excluded firm characteristics using a pseudo poisson estimator with conditional mean specification \eqref{eq:tfp_illustration}; TFP is estimated in levels as $\exp(y_{it} - \hat{\beta}_{1i} k_{it} - \hat{\beta}_{2i} l_{it} - \hat{\beta}_{3i} m_{it})$; pseudo-poisson estimates can be interpreted as semi-elasticities and are relative to firms that do not export or import, pay below median wages and do not advertise; for instance, a firm in industry 311 engaged in international trade is $100 (\exp(0.027) - 1)\% \approx 2.7368\%$ more productive than a firm with the same characteristics that is not engaged.
			\item \emph{Source:} Data taken from the replication package of \textcite{gnr2020}.
		\end{tablenotes}
	\end{threeparttable}
\end{sidewaystable}
I find sizable differences, relative to the magnitude of the standard errors, between the estimated output elasticities for the ex-ante and data-driven classification. There are also some similarities between the estimates for Industry 311 and Group 2 and the estimates for Industry 381 and Group 3. These similarities might be explained by the large overlap in the matching of the firms. The estimated capital elasticities for the data-driven classification are substantially larger than those for the ex-ante classification whereas the ranges of the estimated labor and intermediate input elasticities are wider. The differences in the estimated output elasticities also lead to different conclusions about the heterogeneity in TFP. For Industry 331, I find negative but insignificant TFP differences between firms that engage in international trade and those who do not. The differences in all other industries are positive, but only for industry 321 significantly different from zero.\footnote{This is surprising, as there is wide agreement that firms that engage in international trade, such as exporters, are more productive. Two reasons mentioned by \textcites{bw1997}{bj1999} are self selection into exporting, e.g. because entering foreign markets is costly, and learning by exporting, e.g. knowledge spillovers.} The remaining conclusions about heterogeneity in TFP remain qualitatively the same for both classification approaches. For instance, I find that firms with above median wages are significantly more productive, while firms that advertise are not significantly different from those that do not advertise.

\section{Concluding Remarks}
\label{sec:conclusion}

I propose a fully data-driven estimation procedure for production functions with latent group structures. My approach combines recent identification strategies with the classifier-Lasso of \textcite{ssp2016}. Simulation experiments confirm that my estimation procedure is well suited to deal with latent firm heterogeneity in sufficiently long panels. The practical relevance is illustrated with a panel of Chilean firms.

% Estimator can be extended?
Future research could relax the time-homogeneity assumption of the model parameters within a latent group, e.g. by introducing smooth time-varying model parameters as in \textcite{swj2019}. Another possible extension could be to think of the model parameters not as group-specific fixed constants, but as firm-specific random coefficients with unknown distribution function. Group-specific estimates of the model parameters could then be interpreted as representative points of an unknown distribution.

% References
\clearpage
\printbibliography

% Appendix
%\clearpage
%\appendix
%\section{Appendix}
%\subsection{Simulation Experiments}

%\subsubsection{Representative Firm Model}
%\label{sec:firm_model}
%\clearpage

%\subsubsection{Additional Results}

%\begin{figure}[!htbp]
%	\centering
%	\caption{Classification and Estimation of $\gamma_{j}$}
%	\includegraphics{figure_gamma.pdf}
%	\label{fig:properties_gamma}
%\end{figure}

%\begin{figure}[!htbp]
%	\centering
%	\caption{Classification and Estimation of $\beta_{j}$}
%	\includegraphics{figure_beta.pdf}
%	\label{fig:properties_beta}
%\end{figure}

\end{document}